%% file: c19.tex
\RequirePackage{rotating}
\documentclass[fleqn,usenatbib]{mnras}

\usepackage[T1]{fontenc}
\usepackage{ae,aecompl}

\usepackage{newtxtext}
\usepackage[varvw]{newtxmath}

\usepackage{graphicx}	
\usepackage{amsmath}	
\usepackage[british]{babel}
\usepackage[]{units}
\usepackage{caption}
\usepackage{booktabs}
\usepackage{tabularx}
\usepackage[para,referable]{threeparttablex}
\usepackage{xcolor}
\usepackage{microtype}
\hypersetup{pdfauthor={Errani et al.},
            pdftitle={},
            pdfkeywords={},
            bookmarksnumbered=true}

\newcommand{\rmax}{r_\mathrm{mx}}	%
\newcommand{\Vmax}{V_\mathrm{mx}}	%
\newcommand{\Mmax}{M_\mathrm{mx}}	%
\newcommand{\kms}{\mathrm{km\,s^{-1}}}		%
\newcommand{\Rh}{R_\mathrm{h}}	%

\newcommand{\Msol}{\mathrm{M_{\odot}}}
\newcommand{\Lsol}{\mathrm{L_{\odot}}}

\newcommand{\Vc}{V_\mathrm{c}}

\newcommand{\kpc}{\mathrm{kpc}}
\newcommand{\pc}{\mathrm{pc}}
\newcommand{\Gyrs}{\mathrm{Gyrs}}
\newcommand{\rperi}{r_\mathrm{peri}}

\newcommand{\Torb}{T_\mathrm{orb}}
\newcommand{\rapo}{r_\mathrm{apo}}

\newcommand{\E}{\mathcal{E}}

\newcommand{\sigmalos}{\sigma_\mathrm{los}}
\newcommand{\FeH}{\mathrm{[Fe/H]}}

\title[C-19: Tidal debris of a dark matter-dominated GC?]{The \emph{Pristine} survey -- XVIII. \\C-19: Tidal debris of a dark matter-dominated globular cluster?}

\author[]{Rapha\"el Errani${}^{1}$\thanks{errani@unistra.fr},
Julio F. Navarro${}^2$,
Rodrigo Ibata${}^1$,
Nicolas Martin${}^1$, 
Zhen Yuan${}^1$,
\newauthor
David S. Aguado${}^{3,4}$,
Piercarlo Bonifacio${}^5$,
Elisabetta Caffau${}^5$,
\newauthor
Jonay I. Gonz\'alez Hern\'andez${}^{6,7}$,
Khyati Malhan${}^8$,
Rub\'en S\'anchez-Janssen${}^9$,
\newauthor
Federico Sestito${}^2$,
Else Starkenburg${}^{10}$,
Guillaume F. Thomas${}^{6,7}$,
Kim A. Venn${}^2$
\\
$^1$ Universit\'e de Strasbourg, CNRS, Observatoire astronomique de Strasbourg, UMR 7550, F-67000 Strasbourg, France\\
$^2$ Department of Physics and Astronomy, University of Victoria, Victoria, BC V8P 5C2, Canada\\
$^3$ Dipartimento di Fisica e Astrofisica, Univerisit\`a degli Studi di Firenze, via G. Sansone 1, I-50019 Sesto Fiorentino, Italy\\
$^4$ INAF/Osservatorio Astrofisico di Arcetri, Largo E. Fermi 5, I-50125 Firenze, Italy\\
$^5$ GEPI, Observatoire de Paris, Universit\'e PSL, CNRS, 5 Place Jules Janssen, 92195, Meudon, France\\
$^6$ Instituto de Astrof\'isica de Canarias, E-38205 La Laguna, Tenerife, Spain\\
$^7$ Dpto. Astrof\'isica, Universidad de La Laguna, E-38206 La Laguna, Tenerife, Spain\\
$^8$ Max-Planck-Institut f\"ur Astronomie, K\"onigstuhl 17, D-69117 Heidelberg, Germany\\
$^9$ UK Astronomy Technology Centre, Royal Observatory, Blackford Hill, Edinburgh, EH9 3HJ, UK\\
$^{10}$ Kapteyn Astronomical Institute, University of Groningen, Landleven 12, 9747 AD Groningen, The Netherlands\\
}

\date{Accepted 2022 May 27. Received 2022 May 20; in original form 2022 March 3}

\pubyear{2022}
\volume{}

\begin{document}

\label{firstpage}
\pagerange{\pageref{firstpage}--\pageref{lastpage}} \pubyear{2022}
\maketitle

\begin{abstract}
The recently discovered C-19 stellar stream is a collection of
kinematically associated metal-poor stars in the halo of the Milky
Way lacking an obvious progenitor.  The stream spans an arc of
$\sim 15^\circ$ in the sky, and orbit-fitting suggests an apocentric
distance of $\sim20\,\kpc$ and a pericentre of $\sim10\,\kpc$. The
narrow metallicity dispersion of stars with available spectra,
together with light element abundance variations, suggests a globular
cluster (GC) origin. The observed metallicity ([Fe/H]
$\approx-3.4$), however, is much lower than that of any known GC. In
addition, the width and velocity dispersion of the stream are
similar to those expected from disrupting dwarf galaxies, and
substantially larger than the tidal debris of GCs able to disrupt on
C-19's orbit.  We propose here an unconventional model where the
C-19 progenitor is a dark matter-dominated stellar system with GC-like
abundance patterns. We use N-body simulations to show that the tidal
disruption of a $\sim 100$ pc King-model stellar component embedded
in a $\sim 20$ km/s cuspy cold dark matter halo yields debris
consistent with C-19's observed width and velocity dispersion. The
stellar component of the progenitor is fully disrupted, and is
spread over two distinct streams; one corresponding to C-19 and
another possibly hiding behind the Galactic plane. If such companion
stream were found, it would suggest that dark matter-dominated
dwarfs may also develop GC-like enrichment patterns, a finding that
would inform our theoretical understanding of the formation of
multiple populations in GCs and dwarf galaxies alike.
\end{abstract}




\input{1_introduction}

\input{2_methods}

\input{3_results}

\input{4_conclusions}

\footnotesize{
\section*{Acknowledgements}
RE, RI and ZY acknowledge funding from the European Research Council (ERC) under the European Unions Horizon 2020 research and innovation programme (grant agreement No. 834148). 
DA acknowledges support from the ERC Starting Grant H2020/808240.
JIGH acknowledges financial support from the Spanish Ministry of Science and Innovation (MICINN) project PID2020-117493GB-I00.
ES acknowledges funding through VIDI grant ``Pushing Galactic Archaeology to its limits'' (with project number VI.Vidi.193.093) which is funded by the Dutch Research Council (NWO).
KAV is grateful for funding through the Natural Sciences and Engineering Research Council of Canada (NSERC) Discovery Grants and CREATE programs.
ZY acknowledges support from the French National Research Agency (ANR) funded project ``Pristine'' (ANR-18-CE31-0017). 

\section*{Data availability}
The data underlying this article will be shared on reasonable request to the corresponding author.

\bibliography{c19}
}

\label{lastpage}

\end{document}

%% file: 1_introduction.tex
\section{Introduction}
\label{sec:intro}
The number of known stellar streams in the Milky Way (MW) has increased dramatically over the last few decades thanks to discoveries made by the Sloan Digital Sky Survey \citep[e.g.][]{Odenkirchen2001,Belokurov2006,Koposov2012}, the Dark Energy Survey \citep[e.g.][]{Shipp2018}, and the Gaia satellite \citep[e.g.][]{Malhan2018,Ibata2019,Ibata2021}, among others.

Stellar streams form through tidal stripping of globular clusters (GCs) or dwarf galaxies (DGs) as they orbit in the gravitational potential of the Milky Way. Since globular clusters are in general denser and physically smaller than dwarfs, they produce narrow, kinematically cold stellar streams (like, e.g., the Pal-5 stream, \citealt{Odenkirchen2001}). Dwarfs, on the other hand, are expected to produce thicker and kinematically hotter streams (like, e.g., the massive Sagittarius stream, \citealt{Ibata2001,Newberg2002}). Stream kinematic properties are particularly useful when inferring the progenitor properties of ``orphan'' streams, where no bound remnant of the original system can be identified.

Kinematic signatures, however, are inconclusive on their own. One reason is the recent discovery of ``ultra-faint dwarfs'' (UFDs, usually defined as DGs with $L<10^5\, \Lsol$), whose structural properties (size, velocity dispersion, luminosity) overlap those of some GCs \citep{Simon2019}. Indeed, it often isn't clear whether a system is a GC or a UFD unless a fairly precise estimate of its dynamical mass can be made: objects dominated by dark matter are considered UFDs whereas self-gravitating ones are regarded as GCs.

This litmus test, however, cannot be applied to ``orphan streams'', as they are not self-gravitating systems, so other diagnostics need to be considered. A popular one makes use of the ubiquitous presence of ``multiple populations'' (MPs) in massive, old GCs. These MPs show a narrow dispersion in metallicity (as measured by $\FeH$), but wide variations in the light-element abundances of individual stars \citep{Gratton2004,Gratton2019}. These variations are also often correlated (or anti-correlated) between elements such as C, N, O, Na, Mg, and Al, signalling a complex nucleosynthetic process which operated during GC formation and whose origin has not yet been fully elucidated \citep{BastianLardo2018}.

UFDs, on the other hand, appear to have formed their stars differently, with available data indicating a fairly broad dispersion in $\FeH$ ($\sigma_{\FeH}$) \citep{WillmanStrader2012,Leaman2012}, and no GC-like correlations between light-element abundances \citep{Tolstoy2009,Simon2019}. It is therefore common practice to associate narrow $\sigma_{\FeH}$ and wide variations in light-element abundances with GCs, although even here important exceptions apply: for example, the well-studied $\omega\,$Cen GC has a fairly broad $\sigma_{\FeH}$ \citep{Norris1995}, and younger, more metal-rich GCs appear to lack MPs \citep{Milone2020}, and some UFDs, like Grus-1\footnote{\citet{Ji2019_Gru1} identify Grus-1 as a UFD based on its member stars being deficient in neutron-capture elements \citep[indicative for dwarf galaxies, see, e.g., ][]{Frebel2015}, while \citet{Walker2016_Grus1} refer to Grus-1 more agnostically as an ``ultrafaint stellar system''.}, have unresolved dispersion in the metallicity (and velocity) of the stars measured so far \citep{Walker2016_Grus1, Ji2019_Gru1}.

Stream kinematic properties are also inconclusive as progenitor diagnostics because thin, cold streams originating from GCs may be thickened and ``heated'' if formed in Galactic subhaloes prior to accretion into the Milky Way \citep{Carlberg2018,Carlberg2020,Malhan2021b}. Such heating may also occur because of the gravitational effect of dark matter substructures in the MW halo \citep{Ibata2002heating,Johnston2002,Penarrubia2019}, or because of the dispersal of stream stars in axisymmetric \citep{Erkal2016} as well as non-axisymmetric or time-evolving potentials \citep{Price-Whelan2016}.

The recent detection of C-19, a stream of metal-poor stars recently discovered in the MW halo by combining the power of Gaia proper motions with the photometric metallicities of the Pristine survey, has renewed interest in this topic \citep{Martin2022_C19}. Interestingly, C-19 has kinematic properties that suggest a dwarf galaxy origin \citep[i.e., a width and velocity dispersion of $\approx180\,\pc$ (for a heliocentric distance of $20\,\kpc$) and $\approx6\,\kms$, respectively, see][]{Yuan2022}, but the element abundances of the few stars measured so far indicate a narrow spread in metallicity ($\sigma_\FeH<0.18$, \citealt{Martin2022_C19}) and light-element abundance variations akin to those in GCs. At the same time, at $\FeH\approx-3.4$, C-19 is more metal-poor than any other dwarf galaxy \citep{Chiti2021} or known globular cluster \citep{Beasley2019}, so elucidating the properties of its progenitor is particularly intriguing.

We explore here an unconventional model, where C-19's progenitor is envisioned as a dark matter-dominated dwarf galaxy with GC-like enrichment patterns. As we show below, this model circumvents a major difficulty afflicting GC progenitor models: no known GC can disrupt fully on C-19's orbit and leave a stream as thick and as kinematically hot as observed. Either the stream has been heated after formation by some of the external effects mentioned earlier, or C-19's progenitor was unlike any GC or dwarf system that has survived to the present.

This contribution is meant to explore the consequences of assuming that C-19's progenitor was a dark matter-dominated dwarf, and to identify features of this scenario that may falsify it. Our intention is not to deny the possibility that the C-19 stream may have indeed originated from a conventional GC (with an unconventional dynamical heating history), nor to survey the full range of possible origins for C-19, but rather to explore the possibility that the most metal-poor stellar systems to form in the early Universe might have differed systematically from the examples (GCs or UFDs) that have survived to the present day.  Should C-19's progenitor be actually a dark matter-dominated low-surface-brightness UFD, as proposed here, this would be extremely useful for theoretical models of the origins of MPs in globular clusters, many of which assume that the high density environment of present-day GCs is critical for the existence of MPs.

The paper is structured as follows. We introduce the numerical modelling of the stream in Sec.~\ref{SecModeling}, and discuss our main results in Sec.~\ref{SecResults}. In particular, we explore the evolution of systems modelled after three well-known GCs (NGC 2419, Pal-5 and Pal-14) placed on C-19's current orbit, and discuss whether it is possible to match the properties of C-19 without leaving behind at the same time a luminous bound remnant (Sec.~\ref{sec:results_GC}). We then explore the stripping of a cold dark matter-dominated dwarf galaxy model in Sec.~\ref{sec:results_dwarf}. We end with a brief summary of our main conclusions in Sec.~\ref{sec:conclusions}.

%% file: 2_methods.tex
\section{C-19 modelling}
\label{SecModeling}
\subsection{Observational summary}

\begin{figure}
\centering
\includegraphics[width=8.5cm]{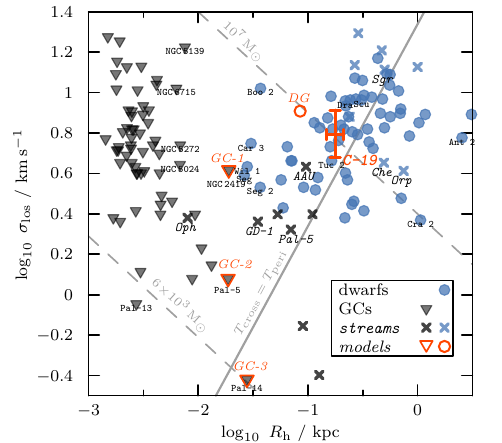}
\caption{Half-light radii $\Rh$ and line-of-sight velocity dispersions $\sigmalos$ of dwarf galaxies (blue, filled circles) and globular clusters (grey, filled triangles) with luminosities $L<10^7\,\Lsol$, compared against the characteristic width and velocity dispersion (see footnotes \ref{footnote:width}, \ref{footnote:dispersion}) of the C-19 stream (red errorbars). Indicative values of width and velocity dispersion are also shown for other stellar streams (crosses, in blue for streams with dwarf galaxy progenitors, and in grey for those with globular cluster progenitors). $N$-body models explored in this work are marked in red. References for the data shown are given in footnote~\ref{footnote:dsph_GC_data}. A line of constant crossing time $T_\mathrm{cross} \sim 2 \pi \Rh/\sigmalos$ that equals the circular orbital time at pericentre $T_\mathrm{peri} \approx 2 \pi \rperi / (240\,\kms)$, is shown in grey. Systems that fall to the right of this line are likely to experience significant tidal stripping on the C-19 orbit. 
The two dashed diagonal lines correspond to self-gravitating Plummer models (Eq.~\ref{eq:GC_mtot}) with fixed total masses of $M_\star = 6\times10^3\,\Msol$ (the minimum stellar mass of C-19, see text), and of $10^7\,\Msol$.
The characteristic width and velocity dispersion of the C-19 stream place it closer to the regime of dwarf galaxies than to the regime of globular clusters.}
\label{fig:dsph_vs_GCs}
\end{figure}

\begin{figure}
\centering
\includegraphics[width=8.5cm]{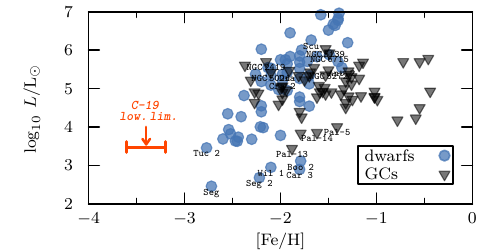}
\caption{Metallicity $\FeH$ and luminosity $L$ of the dwarf galaxies (blue, filled circles) and globular clusters (grey, filled triangles) shown in Fig.~\ref{fig:dsph_vs_GCs}. The C-19 progenitor (shown in red, with errorbars in $\FeH$ corresponding to $\pm0.2\,\mathrm{dex}$ systemic uncertainty) appears unique in its properties when compared against known dwarf galaxies and globular clusters.}
\label{FigLFeH}
\end{figure}

The extreme metallicity of C-19 stars \citep[$\FeH=-3.38\pm0.06 \,\mathrm{(stat.)}\,\pm0.20\,\mathrm{(syst.)}$;][]{Martin2022_C19} is well below the metallicity of the second most metal-poor stream discovered so far, the Phoenix stream \citep[$\FeH\approx-2.7$;][]{Wan2020}, and well below the most metal-poor dwarf galaxy, Tucana~2 \citep[$\FeH\approx-2.8$;][]{Chiti2021}. It is also more metal-poor than all known globular clusters, implying that its progenitor was a unique system by any account.

As mentioned above, its width\footnote {\label{footnote:width}The width of the stream is defined as one standard deviation in right ascension $\alpha$ at fixed declination $\delta$ for a heliocentric distance of $20\,\kpc$, multiplied by $\cos(\delta)$.} and velocity dispersion\footnote{\label{footnote:dispersion} The transverse velocity dispersion is estimated as the dispersion in line-of-sight velocity with respect to the fitted orbit at fixed declination $\delta$.} \citep[$177\pm25\,\pc$ and $6.2^{+2.0}_{-1.4}\,\kms$, respectively, see][]{Yuan2022} suggest a dwarf galaxy origin rather than a globular cluster one. We show this in Figure \ref{fig:dsph_vs_GCs}, where we compare the width and velocity dispersion of C-19 (red symbols with errorbars) with a compilation\footnote
{Data for dwarf galaxies is as compiled in \citet{McConnachie2012} (version January 2021, width updated data for Antlia 2, Crater 2 \citep{Ji2021}, Tucana \citep{Taibi2020}, Tucana 2 \citep{Chiti2021}, And 19 \citep{Collins2020} and And 21 \citep{Collins2021}), and for globular clusters as in \citet{Harris1996} (version December 2010, and updated half-light radii and velocity dispersions for Pal-5 from \citealt{Kuzma2015,Gieles2021}; for NGC 2419 from \citealt{Baumgardt2009}, and for Pal-14 from \citealt{Hiker2006,Jordi2009}). Width and velocity dispersions for stellar streams are from the $S^5$ survey \citep[][table~1]{Li2021}, with stream width defined as one standard deviation of a Gaussian profile. In addition, data is shown for the Sagittarius stream (width and velocity dispersion from \citealt{Ibata2020} and \citealt{Gibbons2017}, respectively), the Palomar-5 stream (\citealt{Bonaca2020}, \citealt{Kuzma2015}), the GD-1 stream (\citealt{Koposov2010}, \citealt{Gialluca2021}), and the C-19 stream (both width and dispersion from \citealt{Yuan2022}, for a heliocentric distance of $20\,\kpc$).
\label{footnote:dsph_GC_data}}
of half-light radii, $\Rh$, and line-of-sight velocity dispersions, $\sigmalos$, of globular clusters (filled grey triangles) and dwarf galaxies (filled blue circles) with $L<10^7\,\Lsol$. The same figure also includes data for other streams in the halo of the Milky Way (blue crosses for dwarf galaxy streams, and grey crosses for GC streams). The globular cluster stream closest in width and velocity dispersion to C-19 is the ``Atlas and Aliqa Uma'' stream (AAU), which is likely perturbed from interaction with the Sagittarius dwarf \citep{Li2021_AAU}.

Fig.~\ref{FigLFeH}, on the other hand, compares the metallicity of C-19 and its inferred luminosity $\sim 3 \times 10^3\, \Lsol$ \citep[more appropriately, its lower limit, as derived by][]{Martin2022_C19} with dwarf galaxies and GCs. Clearly, regardless of its luminosity, the progenitor of C-19 has no known parallel in our Local Group.

Aside from its width and velocity dispersion, another C-19 property useful for comparison with our models is its angular extent. This is estimated to be $\sim 15^\circ$ in the sky,  with only one star candidate beyond that region \citep{Yuan2022}, and no identifiable bound remnant. We shall take this to imply that the detected stream contains the bulk of the stars in the disrupted progenitor (assumed fully disrupted). Since streams grow longer and sparser as they orbit the Galactic halo, this, in turn, suggests that the stream is relatively ``young'', in the sense that its disruption happened relatively recently (i.e., in the most recent past few orbital periods).

Because the stream's pericentric distance of $\rperi\sim 10$ kpc and apocentric distance $r_{\rm apo}\sim 20$ kpc indicate an orbital period of just $T_{\rm orb}\sim 300$ Myr, this implies that either (i) the progenitor has been recently accreted into the MW halo, or that (ii) the stellar component of C-19 was deeply embedded in a more massive (dark) halo that has only recently been sufficiently stripped to affect its stars. Since (i) is disfavoured by the relatively short orbital time and small apocentric distance, we take this observation to support, at least in principle, the premise that C-19's progenitor could have been more massive and extended in the past.

\subsection{Numerical methods}
\label{sec:NumMethods}

We summarize in this section the $N$-body models and numerical tools used in this work. 

\subsubsection{Host galaxy model}

We model the Milky Way using a static, axisymmetric, analytical potential with a circular velocity of $\Vc(R_\odot) = 240\,\kms$ at the solar circle, $R_\odot=8.3\,\kpc$. We use the parametrisation of \citet{EP20}, which in turn is derived from the \citet{McMillan2011} model. 

In this parametrization, the dark matter halo is modelled using a spherical NFW profile with virial\footnote{Virial quantities are computed within spheres of mean density equal to 200 times the critical density for closure, $\rho_{\rm crit}=3H_0^2/8\pi G$, and are indicated with a ``200'' subscript.} mass $M_{200}=1.15 \times 10^{12}\,\Msol$, virial radius $r_{200}=192\,\kpc$ and concentration parameter $c=9.5$. The bulge is modelled using a \citet{hernquist1990} sphere with total mass $M_\mathrm{b} = 2.1 \times 10^{10}\,\Msol$ and scale length $a_\mathrm{b}=1.3\,\kpc$. Thick and thin discs are axisymmetric \citet{miyamoto1975} potentials with masses $M_\mathrm{thick}=2.0 \times 10^{10}\,\Msol$ and $M_\mathrm{thin}=5.9\times10^{10}\,\Msol$, and horizontal and vertical scale lengths $a_\mathrm{thick}=4.4\,\kpc$, $a_\mathrm{thin}=3.9\,\kpc$ and $b_\mathrm{thick}=0.92\,\kpc$, $b_\mathrm{thin}=0.31\,\kpc$, respectively.

\subsubsection{Orbit}
\label{sec:orbit}

We use the potential described above to fit an orbit that matches approximately the sky location  $\{\alpha,\delta\}$, proper motions $\{\mu_\alpha*,\mu_\delta\}$, and radial velocities $v_\mathrm{los}$ (if available), of C-19 member candidates identified by \citet{Yuan2022}. Our orbit fitting assumes a solar motion relative to the local standard of rest as in \citet{SB2010}.

\begin{figure}
 \centering
  \includegraphics[width=8.5cm]{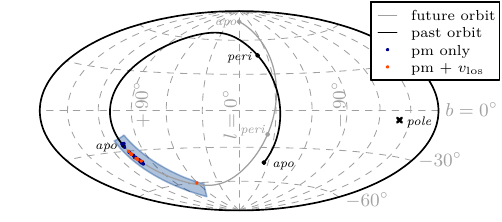}
\caption{Orbit fitted to the C-19 members in heliocentric galactic coordinates $l,b$, integrated backward (black line) and forward (grey line) in time from the apocentre closest to the C-19 stream. Apo- and pericentres are marked along the orbit. Stream members with radial velocity measurements are shown in red, those with proper motion measurements only are shown in blue. The blue shaded area corresponds to the segment with right ascension $345^\circ < \alpha < 360^\circ$ and declination $-15^\circ < \delta < 40^\circ$ shown in Fig.~\ref{fig:gc3_vs_data} and Fig.~\ref{fig:dSph_vs_data}. The current orbital pole ($l\approx-140^\circ,b\approx-6^\circ$) is marked by a cross. }
\label{fig:galactic_orbit}
\end{figure}

\begin{figure}
 \centering
  \includegraphics[width=8.5cm]{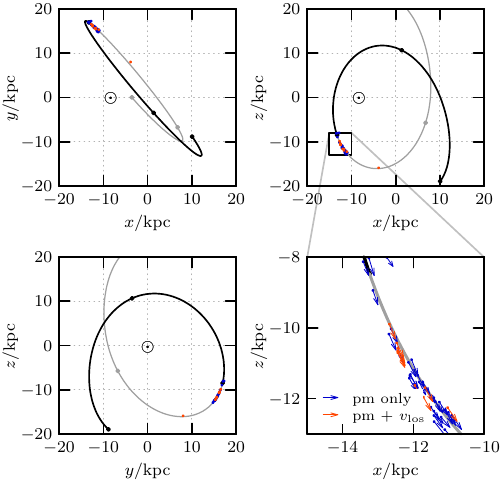}
\caption{Orbit fitted to C-19 member (black solid for integration backward in time, grey for forward integration), like Fig.~\ref{fig:galactic_orbit}, but in galactocentric Cartesian coordinates. Apo- and pericentres are marked along the orbit. The Sun is located at $(x_\odot,y_\odot,z_\odot) = (-R_\odot, 0, 0)$. The bottom right-hand panel shows a zoom-in on the main stream segment, with C-19 member velocities shown as vectors. For those member stars without radial velocity measurement (shown in blue), the radial velocity of the orbit is used in addition to the measured proper motions to compute the velocity vector shown. }
\label{fig:cartesian_orbit}
\end{figure}

The fitted orbit has a pericentre of $\rperi = 11\,\kpc$, an apocentre of $\rapo = 23\,\kpc$, and an orbital period of $\Torb = 0.34\,\Gyrs$. The orbit is shown in Fig.~\ref{fig:galactic_orbit} in heliocentric Galactic coordinates $\{l,b\}$. 
Fig.~\ref{fig:cartesian_orbit} shows the orbit in Galactocentric Cartesian coordinates, with the Sun at $(x_\odot,y_\odot,z_\odot) = (-R_\odot, 0, 0)$.

The bottom right-hand panel of Fig.~\ref{fig:cartesian_orbit} shows a zoom-in on the main stream segment, with velocities shown as vectors. For member stars without measured radial velocities (shown in blue), the radial velocity of the orbit at their location is used to obtain the velocity vectors in Cartesian coordinates.

\subsection{Globular cluster models}
\label{sec:GC_model}
We model globular clusters as self-gravitating Plummer spheres with isotropic velocity dispersion,
\begin{equation}
 \rho_\mathrm{P}(r) = \frac{3M_\star}{4\pi a^3}  \left[1+\left( {r}/{a} \right)^2 \right]^{-5/2} 
\end{equation}
with total mass $M_\star$, and scale radius $a$. 
For spherical Plummer models, the projected half-light radius coincides with scale radius, $\Rh = a$.

The line-of-sight velocity dispersion profile equals \citep[see, e.g.,][equ.~43]{Dejonghe1987}
\begin{equation}
\label{equ:Plummer_sigmalos}
 \sigmalos^2(R) = \frac{GM_\star}{a} \frac{3 \pi}{64} \left[ 1+\left( {R}/{a} \right)^2 \right]^{-1/2}~,
\end{equation} 
which implies that the total mass, $M_\star$, may be estimated from the projected half-light radius, $\Rh$, and the central line-of-sight velocity dispersion, $\sigmalos(0)$,
\begin{equation}
\label{eq:GC_mtot}
M_\star \approx 6.8\, R_\mathrm{h} \sigmalos^2(0) / G ~.
\end{equation}

The importance of Galactic tides on the evolution of a cluster may be gauged by comparing its inner crossing time, $T_\mathrm{cross}\sim 2 \pi R_{\rm h}/\sigmalos(0)$, with the circular orbital time at pericentre,  $T_\mathrm{peri} \approx 2 \pi \rperi/ (240\,\kms)$.

The latter is shown in Fig.~\ref{fig:dsph_vs_GCs} as a solid grey line. Systems on or to the right of this line are expected to be severely affected by tides on such orbit; systems well to the left should be able to survive relatively unscathed for many orbital periods.

As may be seen from Fig.~\ref{fig:dsph_vs_GCs}, most GCs are too dense to disrupt on an orbit like that of C-19, with $r_{\rm peri} =11$ kpc; only Pal-14, the least dense known MW GC, is likely to disrupt fully on such orbit. With this caveat, we explore three different self-gravitating GC models, with parameters chosen as \emph{GC-1}, \emph{GC-2}, and \emph{GC-3} in  Fig.~\ref{fig:dsph_vs_GCs}. These models span a fairly wide range of velocity dispersions and radii, and hence also of stellar mass, as listed in Table~\ref{tab:GC_param}. Model \emph{GC-1} is a fairly dense and massive ($M_\star \approx 5.1 \times 10^5 \, \Msol$) GC, with parameters resembling those of NGC 2419. Model \emph{GC-3}, on the other hand is a fluffy, low mass ($M_\star \approx 6 \times 10^3 \, \Msol$) cluster resembling Pal-14.
Model \emph{GC-2} has intermediate properties, resembling those of Pal-5.

The $N$-body realizations have $N=10^6$ particles each and are generated using the code of \citet{EP20}, available online\footnote{\url{https://github.com/rerrani/nbopy}}.

\begin{table}
\centering
\begin{threeparttable}
\caption{Projected half-light radii $\Rh$, line-of-sight velocity dispersions $\sigmalos$ and total masses $M_\star$ of the globular cluster models used in this work. Masses are estimated using Equation~\ref{eq:GC_mtot}.}
\label{tab:GC_param}
\begin{tabularx}{\linewidth}{ccccc}
\toprule
Model    & Equivalent GC     & $\Rh/\pc$              & $\sigmalos/\kms$               & $M_\star/\Msol$        \\ \midrule
\emph{GC-1}  & NGC 2419          & $19.0$\tnotex{tn:1}    &  $4.14 \pm 0.48$\tnotex{tn:1}  & $5.1 \times 10^5 $ \\
\emph{GC-2} & Pal-5             & $18.7$\tnotex{tn:2}    &  $1.2 \pm 0.3$\tnotex{tn:3}    & $4.2 \times 10^4 $ \\ 
\emph{GC-3}& Pal-14            & $27.8$\tnotex{tn:4}    &  $0.38 \pm 0.12$\tnotex{tn:5}  & $6.3 \times 10^3 $ \\
\bottomrule
\end{tabularx}
\begin{tablenotes}
 \item[1] \label{tn:1} \citet{Baumgardt2009}
 \item[2] \label{tn:2} \citet{Gieles2021}
 \item[3] \label{tn:3} \citet{Kuzma2015}
 \item[4] \label{tn:4} \citet{Hiker2006} 
 \item[5] \label{tn:5} \citet{Jordi2009}
\end{tablenotes}

\end{threeparttable}
\end{table}

\begin{figure}
 \centering
  \includegraphics[width=8.5cm]{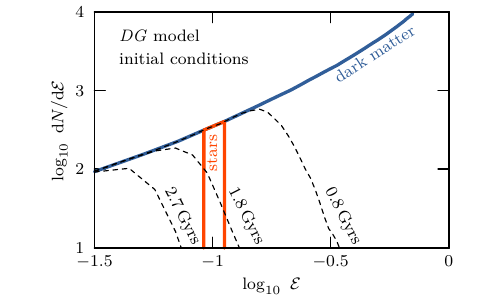}
 
 \includegraphics[width=8.5cm]{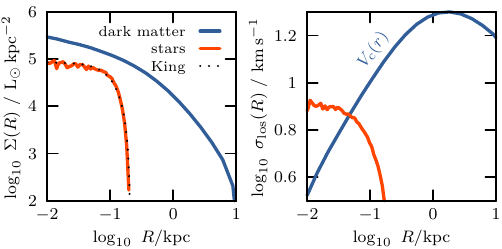}

\caption{Initial conditions of the dwarf galaxy-like model. Dark matter properties are shown in blue, those of stars in red. The total stellar luminosity is scaled to $3 \times 10^3\,\Lsol$. The top panel shows the initial energy distributions of dark matter and stars, as well as the energy distribution of those particles which remain bound after $0.8\,\Gyrs$, $1.8\,\Gyrs$ and $2.7\,\Gyrs$, corresponding to the snapshots shown in Fig.~\ref{fig:dwarf_overview_large}. The initial (2D) surface brightness profile is shown in the bottom left, together with the projected initial dark matter density (arbitrary normalisation). The stellar profile is cored, with a projected half-light radius of $\Rh \approx 85\,\pc$. A \citet{King1962} surface brightness profile with concentration parameter $c_\mathrm{K}=1$ and King core radius $R_\mathrm{K}=210\,\pc$ is shown, for reference, as a black dotted curve. The stellar line-of-sight velocity dispersion profile is shown in the bottom right, with a central line-of-sight velocity of $\sigmalos \approx 8.1\,\kms$.  }
\label{fig:progenitor_ICs}
\end{figure}

\subsection{Dark matter-dominated dwarf galaxy-like model}
\label{sec:dwarf_model}

Our dwarf galaxy-like model assumes that the stellar component of C-19's progenitor may be modelled as spherical King-like models \citep{King1962} embedded in a Navarro-Frenk-White \citep[hereafter, NFW,][]{Navarro1996a,Navarro1997} halo. Stars are assumed to be subdominant gravitationally, and can therefore be modelled as massless tracers of the NFW halo, as detailed below. 

The NFW dark matter subhalo model is given by \begin{equation}
 \rho_\mathrm{NFW} = \rho_\mathrm{s} \left(r/r_\mathrm{s}\right)^{-1} \left(1 + r/r_\mathrm{s}\right)^{-2}
\end{equation}
where $\rho_0$ and $r_\mathrm{s}$ denote the scale density- and radius, respectively. The corresponding circular velocity curve $\Vc = \sqrt{GM(<r)/r}$ reaches its maximum of $\Vmax \approx 1.65\,r_\mathrm{s}\sqrt{G\rho_\mathrm{s}}$ at a radius $\rmax \approx 2.16\,r_\mathrm{s}$. 
The $N$-body realization consists of $10^7$ particles. The NFW density profile is exponentially tapered beyond $10\,r_\mathrm{s}$ to prevent the divergence of the total mass.

We choose the virial mass of the subhalo to approximately correspond to the minimum halo mass able to form stars imposed by the hydrogen cooling limit (HCL) at an early redshift, chosen here to be $z=5$ \citep{Benitez-Llambay2019}. The motivation for this choice is mainly one of simplicity: given C-19's extreme metallicity, its progenitor is likely one of the oldest systems to form in the Universe, and therefore, if dark matter dominated, it should populate some of the lowest mass systems where efficient energy dissipation of the collapse energy may occur. More specifically, we chose a maximum circular velocity of $\Vmax=20\,\kms$, and a radius of maximum circular velocity $\rmax=1.7\,\kpc$ in order to match the $z=5$ mean mass-concentration relation \citep{Ludlow2016}. The mass enclosed within $\rmax$ equals $\Mmax \approx 1.6\times10^8\,\Msol$ and the virial mass of the subhalo is $M_{200} \approx 3 \times 10^8\, \Msol$.

Stars are embedded in this subhalo as massless tracers, and can therefore be modelled by attaching to each dark matter particle a ``stellar probability'' that depends on their initial binding energy. We express the binding energy in dimensionless units,
\begin{equation}
\label{equ:energy}
 \E = 1 - E/\Phi_0~,
\end{equation}
where $\Phi_0\equiv \Phi(r=0)$ corresponds to the potential minimum of the dark matter halo. In these units, the most-bound state is $\E = 0$, and $\E = 1$ corresponds to the boundary between bound and unbound. 
For NFW profiles, the potential minimum is directly related to the maximum circular velocity, $\Phi_{0,\mathrm{NFW}} = - 4 \pi G \rho_\mathrm{s} r_\mathrm{s}^2 \approx -4.64\, \Vmax^2$ \citep[for details, see, e.g.,][]{ENIP2021}. For the tapered NFW models used in this work, we have $\Phi_0 \approx -4.32\, \Vmax^2$. 

We select the stellar energies so that the stars form a deeply embedded cored profile within the NFW halo dark matter cusp, with a (2D) half-light radius of $\Rh \approx 85\,\pc$, and a central line-of-sight velocity dispersion of $\sigmalos \approx 8.1\,\kms$. We accomplish this by simply assigning a uniform ``stellar probability'' to all dark matter particles between a specified minimum, $\E_{\rm min}$, and maximum, $\E_{\rm max}$, initial binding energy (see top panel of Fig.~\ref{fig:progenitor_ICs}). Probabilities are set to zero for dark matter particles outside that range.

This choice results in a stellar profile well described by a sharply truncated \citet{King1962} surface brightness profile with concentration parameter $c_\mathrm{K}=1$ and King tidal radius $R_\mathrm{t,K}=210\,\pc$.
(See Table~\ref{tab:dwarf_model} for a summary of the model parameters.)

Figure~\ref{fig:progenitor_ICs} shows the resulting stellar density and velocity dispersion profiles (shown in red) compared with the dark matter (in blue). Note that because the stars are assumed to be subdominant, the stellar mass of the galaxy can be rescaled arbitrarily. Our dwarf galaxy-like model for the C-19 progenitor assumes a total stellar luminosity consistent with the minimum inferred from the minimum integrated luminosity of C-19, i.e., $3 \times 10^3 \, \Lsol$ \citep{Martin2022_C19}. Using a mass-to-light ratio for old stellar populations of $\sim2$ \citep{Maraston05,Woo2008}, this corresponds to a stellar mass of $\sim 6 \times 10^3\,\Msol$.

\begin{table}
\centering
\caption{Properties of the dark matter subhalo model used in this work. The maximum circular velocity $\Vmax$ is chosen to approximately correspond to the minimum halo mass imposed by the hydrogen cooling limit \citep{Benitez-Llambay2019}. The radius of maximum circular velocity $\rmax$ matches the $z=5$ mean mass-concentration relation \citep{Ludlow2016}. Half-light radius and line-of-sight velocity dispersion correspond to the mass-less stellar tracer with energies $0.08 \leq \E_\star 0.12$ as shown in Fig.~\ref{fig:progenitor_ICs}. }
\label{tab:dwarf_model}
\begin{tabular}{c@{\hskip 0.3cm}c@{\hskip 0.3cm}c@{\hskip 0.3cm}c@{\hskip 0.3cm}c@{\hskip 0.3cm}c}
\toprule
Model & $\rmax/\kpc$ & $\Vmax/\kms$     & $\Mmax/\Msol$        & $\Rh/\kpc$ & $\sigmalos/\kms$  \\ \midrule
\emph{DG}& $1.7$        &  $20.0$          & $1.6 \times 10^8 $   & 0.085       & 8.1               \\ 
\bottomrule
\end{tabular}
\end{table}

\subsection{N-body code}

We use the particle-mesh code \textsc{superbox} \citep{Fellhauer2000} to evolve the N-body particles of the GC and DG models in the specified Milky Way potential. The code uses two comoving grids, centred on the $N$-body model, with $128^3$ grid cells each.  The sizes of the co-moving grids are chosen relative to the size of the $N$-body model, resulting for Plummer (NFW) models in cell spacings of $\Delta x \approx 10\,a / 128$ ($\Delta x \approx 10\,\rmax/128$) for the medium resolution grid and $\Delta x \approx a / 128$ ($\Delta x \approx \rmax / 128$) for the high resolution grid.

The equations of motions are solved using a leap-frog-scheme with a time step $\Delta t = T_\mathrm{a}/200$ for Plummer models, where $T_\mathrm{a}$ denotes the period of a circular orbit at radius $a$, and with a time step $\Delta t = T_\mathrm{mx}/400$ for NFW models, where $T_\mathrm{mx}$ denotes the period of a circular orbit at $\rmax$. Using these choices, a circular orbit with a radius equal to the grid resolution $r=\Delta x$ is resolved, in the case of Plummer models, with $\approx 120$ time steps, and in the case of NFW models, with $\approx 16$ steps.

%% file: 3_results.tex
\section{Results}
\label{SecResults}

We begin by analysing the streams produced by the C-19 globular cluster progenitors described in Section~\ref{sec:GC_model}, before considering dark matter-dominated progenitors in the following subsection.

\subsection{Self-gravitating GC model progenitors}
\label{sec:results_GC}

\begin{figure}
  \centering
  \includegraphics[width=8.5cm]{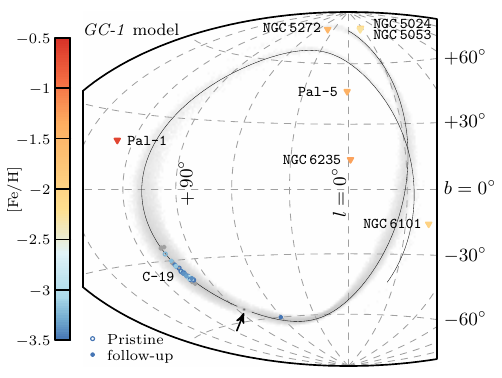}
  \caption{NGC 2419-like model \emph{GC-1}, evolved for $10\,\Gyrs$, in heliocentric Galactic coordinates. The position of the cluster remnant is indicated by an arrow. Shown also are Milky Way GCs \citep{Baumgardt2019} with apocentric- and pericentric distances roughly compatible with C-19. No GC lies on the C-19 orbit (black curve), and all GCs are considerably more metal-rich than C-19 (see colour-coding).}
  \label{fig:no_gc_on_orbit}
\end{figure}

\begin{figure}
 \centering
  \includegraphics[width=6.5cm]{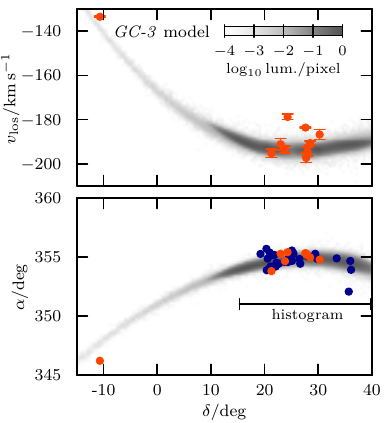}\includegraphics[width=2.0cm]{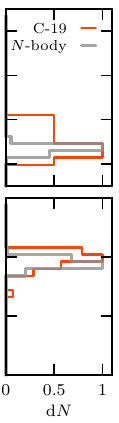}
\caption{Globular cluster model \emph{GC-3} with a Pal-14-like progenitor on the C-19 orbit, evolved for $10\,Gyrs$. The top panel compares the line-of-sight velocities of the $N$-body model as a function of declination $\delta$ against the observational data (red points with errorbars). A histogram, summing over all stars with declination $15^\circ < \delta < 40^\circ$, is shown on the right, comparing the C-19 data (red) against the $N$-body model (grey). The bottom panel compares right ascension $\alpha$ as a function of declination $\delta$ (red and blue points correspond to member stars with and without line-of-sight velocity measurements, respectively).
The model completely disrupts, and after $5\,\Gyrs$ roughly matches the C-19 stream in length and width. However, the velocity dispersion of the simulated stream is much lower than that of the C-19 member stars.}
\label{fig:gc3_vs_data}
\end{figure}

\begin{figure*}
 \centering
  \includegraphics[width=18.8cm]{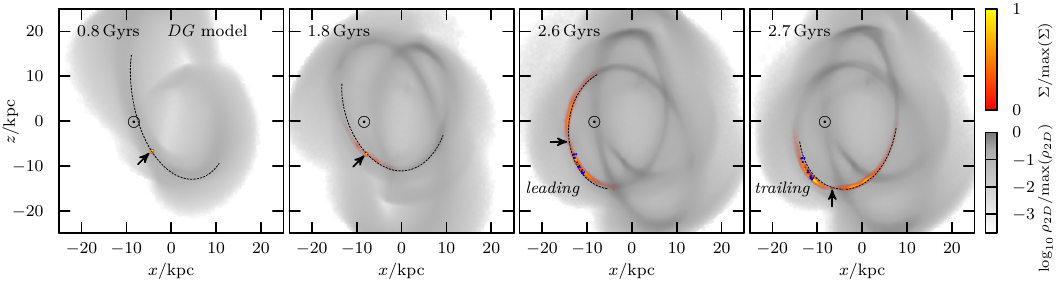}
\caption{Distribution of dark matter (grey) and stars (colour) for the dwarf galaxy-like model at different times after accretion, in galactocentric Cartesian coordinates. 
At $T=0.8\,\Gyrs$ (left-hand panel), some dark matter has been stripped, but the stellar component is largely unaffected by tides. At $T=1.8\,\Gyrs$ (second panel from left), a stellar stream has formed, but a bound stellar cluster remains centred on the remnant dark matter subhalo. At $T=2.6\,\Gyrs$ (third panel from left), the stellar cluster has been completely dispersed, and no bound stars remain within the dark matter subhalo. This scenario is a possible model for the formation of the C-19 stream, with the leading stream corresponding to C-19. Observed stream members are shown using red and blue points (with distances adjusted), in good agreement with the leading stream of the disrupted dwarf galaxy-like model. Note that while no bound stellar remnant remains, the dark matter subhalo does not fully disrupt. The right-hand panel ($T=2.7\,\Gyrs$) shows a similar situation, but here, the trailing tidal stream matches the C-19 data. The centre of mass of the bound dark matter remnant is shown using an arrow in all panels. A video of this simulation is available on the journal website.}
\label{fig:dwarf_overview_large}
\end{figure*}

\subsubsection{NGC 2419-like progenitor (GC-1)}
The \emph{GC-1} model represents a globular cluster with structural properties similar to that of NGC 2419. It has an initial velocity dispersion ($\sim 4\,\kms$) slightly below  that measured for the C-19 stream ($\sim 6\,\kms$), but it is much more massive ($M_\star \approx 5\times 10^5\, \Msol$) than the lower limit on the stellar mass inferred for C-19 ($M_\star \gtrsim 6 \times 10^3\,\Msol$). More importantly, NGC 2419 has a half-light radius of only $\sim 19\,\pc$, implying a relatively high density compared to the inner mean density of the host Galaxy model at its pericentric radius (i.e., it lies well to the left of the grey line in Fig.~\ref{fig:dsph_vs_GCs}).

Because of this, \emph{GC-1} is barely affected by Galactic tides, losing only $15\%$ of its stars over $10\,\Gyrs$ of evolution. This model would predict a massive bound remnant on the same orbit as C-19, at odds with observations, where no known cluster has been found to share C-19's orbit. We show this in Fig.~\ref{fig:no_gc_on_orbit}, which shows in grey stars from the GC-1 progenitor in heliocentric Galactic coordinates $l,b$. The location of the bound luminous remnant (with stellar mass $\sim 4 \times 10^5 \, \Msol$) is indicated by an arrow.

The figure also shows the positions of all known Milky Way GCs with apocentric and pericentric distances in a fairly wide range around the corresponding C-19 values \citep[i.e., $10\,\kpc < \rapo < 30\,\kpc$ and pericentric distances $5\,\kpc < \rperi < 20\,\kpc$, taken from ][]{Baumgardt2019}. 
No GC shown in Fig.~\ref{fig:no_gc_on_orbit} lies closer than $4^\circ$ from C-19's current orbital path. 
In addition, as mentioned earlier, the metallicity of all known GCs is far in excess of that measure for C-19 members, making them incompatible with being potential progenitors to the C-19 stream. We conclude that GC-1 is not viable as a possible progenitor of the C-19 stream.

\subsubsection{Pal 5-like progenitor (GC-2)}

The \emph{GC-2} model has properties similar to those of Pal-5, a relatively fluffy Milky Way globular cluster, substantially less dense and less massive than NGC 2419. As may be seen from Fig.~\ref{fig:dsph_vs_GCs}, this cluster model still has a characteristic crossing time $T_\mathrm{cross} \sim 2 \pi \Rh / \sigmalos$ shorter than the circular orbital time $T_\mathrm{peri} \approx 2\pi \rperi / (240\,\kms)$ at pericentre, and, as a result, we would not expect \emph{GC-2} to be fully disrupted by tides.

Indeed, our results show that \emph{GC-2} loses $\sim 65$ per cent of its initial mass after completing $\sim 30$ orbits in about $10\,\Gyrs$. This implies that \emph{GC-2} would predict that a bound remnant with total luminosity $\sim 10^4\,\Lsol$ should remain visible at present, also at odds with observations. We hence conclude that, just like for GC-1, the GC-2 model is not a viable progenitor of C-19.

\subsubsection{Pal 14-like progenitor (GC-3)}

Finally, we consider a much lower density GC model (\emph{GC-3}), inspired by Pal-14, the least dense GC known in the Galaxy. When placed on C-19's orbit, this system is tidally destroyed after about a dozen of orbits. In order to match C-19's observed length and width, we are forced to study the remnant after just $\sim 5$ Gyrs of evolution. The comparison with C-19, shown in Fig.~\ref{fig:gc3_vs_data}, shows that \emph{GC-3} is able to approximately  match the width (and length) of the observed stream without leaving behind a bound luminous remnant. However, the velocity dispersion of this stream ($\sim 1.4\,\kms$) is much lower than observed for C-19 ($\sim 6\,\kms$), even if analysed just after disruption.
Since streams generally lengthen and become kinematically colder as they evolve \citep[see; e.g.,][]{HelmiWhite1999} it is unlikely that this model can reproduce C-19 even if analysed at a later time.
While ``fanning'' on polar orbits in axisymmetric potentials may increase the width of streams over time \citep[see; e.g.,][]{Erkal2016}, even after $10\,\Gyrs$ of evolution, the velocity dispersion of the \emph{GC-3} stream remains below that of C-19. Earlier accretion times are in conflict with the observed length of the C-19 stream: Evolving the \emph{GC-3} model for $\sim 10\,\Gyrs$ results in a stream more than twice as long as C-19.
We conclude that this model does not provide a good approximation to C-19's data either.

\subsection{Dark matter-dominated dwarf galaxy-like model (\emph{DG})}
\label{sec:results_dwarf}

The failure of the GC progenitors studied above to reproduce C-19 is not entirely unexpected, and may be traced to the assumption that these systems were initially self-gravitating. Indeed, GC progenitors with initial stellar mass similar to that inferred for C-19 must lie close to the grey dashed $6\times10^3\,\Msol$-line in Fig.~\ref{fig:dsph_vs_GCs}.
To disrupt fully on C-19's orbit, GC progenitors must also lie to the right of the solid grey line indicating the circular orbit time at a pericentre of $11\,\kpc$.

These restrictions put potential C-19 progenitors quite far from C-19's locus in the half-light radius vs velocity dispersion plane. Indeed, tentatively\footnote{Eq.~\ref{eq:GC_mtot} was derived for self-gravitating Plummer spheres. Here, we apply it tentatively to the width and velocity dispersion of the C-19 stream, assuming that these values are a rough proxy for size and velocity dispersion of the C-19 progenitor at the time when the stream formed.} applying Eq.~\ref{eq:GC_mtot} to the measured width and dispersion of the C-19 stream, suggests a dynamical mass of $\sim 10^7 \, \Msol$ (see dashed grey line in Fig.~\ref{fig:dsph_vs_GCs}). 

This mass corresponds roughly to that enclosed within $\sim 200\,\pc$ of a halo at the HCL boundary (see Sec.~\ref{sec:dwarf_model}), a fact that motivates the choice of stellar component parameters described in Sec.~\ref{sec:dwarf_model}, which are shown by the red symbol labelled ``\emph{DG}'' in Fig.~\ref{fig:dsph_vs_GCs}.
Dark matter-dominated models have an advantage over self-gravitating GC models, for they allow us to construct systems where the stellar component may fully disrupt, leaving behind no self-bound stellar remnant, even if a small dark matter clump may remain. We explore these issues in more detail below. 

As an illustration, we choose here to simulate the tidal evolution of the dwarf-like model for roughly eight pericentric passages, or $\sim 3\,\Gyrs$. 
Tides strip mass gradually and  ``outside-in'' in terms of initial binding energy \citep[][]{ENIP2021}. This is shown in the top panel of Fig.~\ref{fig:progenitor_ICs}, where thin dashed lines outline the sharp energy truncation experienced by the halo at various stages of evolution (i.e., after $0.8$, $1.8$, and $2.7$ Gyr, respectively). Note that after $2.7$ Gyr of evolution there is no bound stellar remnant left, although a dense clump of dark matter (consisting of the innermost initial dark matter cusp) remains bound \citep[see, also,][]{EN21}.

Figure~\ref{fig:dwarf_overview_large} shows projections of the dwarf-like system at those stages of the evolution.
The dark matter surface density $\rho_\mathrm{2D}$ is shown in grey, and stellar surface brightness $\Sigma$ in colour.
At $t=0.8\,\Gyrs$ (or after $2$ pericentric passages, left-hand panel), some dark matter has been stripped, but the stellar component is still largely unaffected by tides.  At $t=1.8\,\Gyrs$ (second panel from left), a stellar stream has formed, and a bound stellar remnant remains centred on the dark matter subhalo (marked by the arrows in each panel). At $t=2.6\,\Gyrs$ and $t=2.7\,\Gyrs$  (third panel from left, and right-hand panel), the stellar component has been completely dispersed, and no bound stellar progenitor is detectable, consistent with observations.

\begin{figure}
 \centering
  \includegraphics[width=6.5cm]{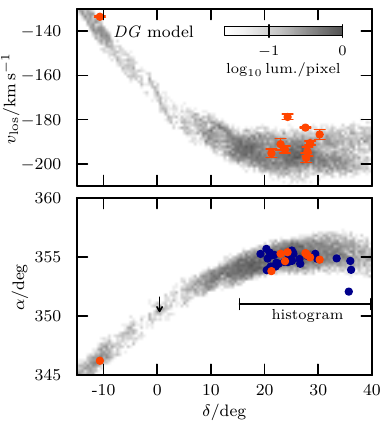}\includegraphics[width=2.0cm]{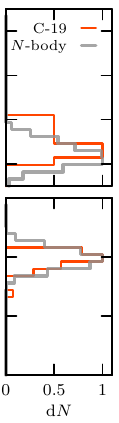}
\caption{Comparison of the dwarf galaxy-like model against velocities (top panel) and positions (bottom panel) of C-19 member stars (as in Fig.~\ref{fig:gc3_vs_data}). The width and dispersion of the $N$-body model match that of the C-19 stream members. The arrow shows the position of the centre of mass of the bound dark matter remnant. }
\label{fig:dSph_vs_data}
\end{figure}

This scenario leads to the formation of two fairly symmetric streams, each of which, at $t=2.6\,\Gyrs$ and $t=2.7\,\Gyrs$, respectively, matches fairly well the width and velocity dispersion of the C-19 stream. This is illustrated in Fig.~\ref{fig:dSph_vs_data} for the trailing stream at $t=2.7\,\Gyrs$; the top panel compares line-of-sight velocities of the simulation model against C-19 member stars, while the bottom panel shows the positions in the sky. Adjacent histograms sum over the main stream segment, defined as all stars with $15^\circ \leq \delta \leq 40^\circ$.

What about the other stream? In the case where C-19 is identified with the trailing stream (i.e., $t=2.7$ Gyr in Fig.~\ref{fig:dwarf_overview_large}) this would predict the presence of a second, nearly symmetric stream segment that is currently not observed. However, if C-19 were identified with the leading stream (i.e., at $t=2.6$ Gyr), then the bulk  of the trailing stream would be at very low Galactic latitudes, and could have easily escaped detection. At least from the kinematic perspective, this scenario provides an attractive explanation for the unexpected width and velocity dispersion of C-19. 

The times quoted here are for illustrative purpose only: matching models with earlier (or later) accretion time may be constructed through small modifications to the initial stellar binding energies. As the angular separation between leading and trailing stream increases over time, the discovery of a second companion stream to C-19, sharing its orbital pole\footnote{While the pole of the C-19 orbit precesses over time, the angular separation between the orbital poles of leading and trailing stream grows at a much lower pace.} ($l\approx -140^\circ, b\approx -6^\circ$, see Fig.~\ref{fig:galactic_orbit}), might allow to constrain the time of accretion of the C-19 progenitor onto the Milky Way. If the C-19 progenitor is still in the process of disrupting, leading and trailing stream may appear as a single, connected structure.

%% file: 4_conclusions.tex
\section{Summary and conclusions}
\label{sec:conclusions}

The C-19 stellar stream is the most metal-poor stream known to date, with a mean metallicity of $\FeH \approx -3.4$. As only a very small spread in metallicity ($\sigma_\FeH\lesssim 0.18$) is observed, it is tempting to associate C-19 with a tidally disrupted globular cluster. However, the unusual width and velocity dispersion of the stream suggest a possible dwarf galaxy origin for the stream. We have explored here scenarios where the progenitor of C-19 is either a self-gravitating globular cluster, or a dark matter dominated stellar system similar in structure to dwarf spheroidal galaxies.

Our main conclusions may be summarised as follows.

\begin{itemize}
 \item[i)] The C-19 stellar stream has a width and a velocity dispersion exceeding those of known globular cluster streams, and structurally place C-19 closer to the regime of dwarf galaxies than that of globular cluster streams. 
 \item[ii)] Self-gravitating globular cluster progenitors similar in structure to NGC 2419 and Pal-5 are too dense to be fully disrupted  on C-19's orbit over a period of $10\,\Gyrs$. These may be ruled out by the lack of a bound luminous remnant associated with C-19.
 \item[iii)] A self-gravitating progenitor similar to the Pal-14 globular cluster matches the integrated luminosity of C-19 and disrupts fully, but yields a stream too kinematically cold to match C-19's observed velocity dispersion.
 \item[iv)] A dark matter-dominated dwarf galaxy-like progenitor, where stars follow a King-like profile embedded in a low-mass cuspy cold dark matter model near the hydrogen cooing limit, is able to reproduce the width and velocity dispersion of C-19, leaving behind no bound stellar remnant.
 \item[v)] The dwarf galaxy progenitor model predicts the existence of a second tidal stream, similar to C-19 but at Galactic latitudes likely coinciding with the Galactic plane, hindering its discovery. The discovery of such a stream, symmetric in structure and luminosity to C-19, would strongly favour the dwarf-like progenitor model. 
 \end{itemize}

Although the dynamical evidence seems to favour a DG-like progenitor for C-19, it is difficult to completely rule out GC-like progenitors. Indeed, different dynamical processes have been discussed in the literature that could increase the width and velocity dispersion of globular cluster streams beyond the values expected from stripping in smooth tidal fields such as the one considered here. 
These include the gravitational effect of (dark) matter clumps in the MW halo \citep{Ibata2002heating,Johnston2002,Penarrubia2019}, the dispersal of stream stars in non-axisymmetric or time-evolving potentials \citep{Price-Whelan2016}, tidal evolution of the GC within a subhalo prior to accretion onto the MW \citep{Carlberg2018,Carlberg2020,Malhan2021b}, as well as collisional processes within the GC that might facilitate tidal disruption \citep{Gieles2021}.
What observations could help us distinguish a globular cluster stream that has been thickened and ``heated''  from the dwarf galaxy-like progenitor we explored here? A non-exhaustive list of diagnostics include the following:
\begin{itemize}
 \item[i)] Signs of collisional processes in the progenitor, like mass-segregation of C-19 member stars along the orbit, might give dynamical hints on the progenitor dynamics and allow to distinguish a (collisional) GC from a (collision-less) dwarf-like system. 
 \item[ii)] Our dwarf galaxy-progenitor model predicts two rather symmetric tidal streams. The detection of a second stream segment, with a luminosity and structure similar to that of the current C-19 segment, would be an indicator in favour of the dwarf-like model, in contrast to the more irregular streams of pre-accretion stripped globular clusters \citep[see, e.g.,][]{Malhan2021b}.
 \item[iii)] A better understanding of the origin of light-element abundance patterns in GCs and in low-metallicity UFDs, together with improved observational constraints on a larger number of C-19 stars, might allow to use them as discriminating diagnostics of C-19's progenitor. Is it actually possible for dwarf galaxies to develop GC-like light-element enrichment patterns while keeping a narrow spread in metallicity? 
\end{itemize}
If the hypothesis of a dark matter-dominated progenitor holds, then C-19 is allowing a glimpse at galaxy evolution at the faintest end of the galaxy luminosity function, and would suggest that the earliest, most metal-poor stellar systems formed may differ substantially from the more metal-rich globulars and dwarfs that have survived to the present day.